# Interfacial Fe segregation and its influence on magnetic properties of CoFeB/MgFeO multilayers

Tomohiro Ichinose,[1,*] Tatsuya Yamamoto,[1] Takayuki Nozaki,[1] Kay Yakushiji,[1] Shingo Tamaru,[1] Shinji Yuasa[1]

1. National Institute of Advanced Industrial Science and Technology (AIST), Research Center for Emerging Computing Technologies, Tsukuba, Japan.

*Corresponding author

e-mail: tomohiro.ichinose@aist.go.jp

*Abstract*

We investigated the effect of Fe segregated from partially Fe-substituted MgO (MgFeO) on the magnetic properties of CoFeB/MgFeO multilayers. X-ray photoelectron spectroscopy (XPS) as well as magnetic measurements revealed that the segregated Fe was reduced to metal and exhibited ferromagnetism at the CoFeB/MgFeO interface. The CoFeB/MgFeO multilayer showed more than 2-fold enhancement in perpendicular magnetic anisotropy (PMA) energy density compared with a standard CoFeB/MgO multilayer. The PMA energy density was further enhanced by inserting an ultrathin MgO layer in between CoFeB and MgFeO layers. Ferromagnetic resonance measurement also revealed a remarkable reduction of magnetic damping in the CoFeB/MgFeO multilayers.

*Introduction*

Magnetoresistive random-access memory (MRAM) is a promising non-volatile memory device which potentially reduces power consumption for data processing. [1-4] State-of-the-art MRAM stores data in magnetic tunnel junctions (MTJs) consisting of a perpendicularly magnetized recording layer. Because the data readout is done via the tunnel magnetoresistance (TMR) effect, high TMR ratio is one of the key parameters in the MTJs. [5-7] On the other hand, the data retention capability of the MTJs is characterized by the thermal stability factor $\Delta = K_\mathrm{u}V/k_\mathrm{B}T$, [1-4] where $K_\mathrm{u}$, $V$, $k_\mathrm{B}$, and $T$ are energy density of perpendicular magnetic anisotropy (PMA), volume of the recording layer, Boltzmann constant, and temperature, respectively. A higher $K_\mathrm{u}$ is strongly demanded for increasing memory density (*i.e.*, reducing $V$) while maintaining a high $\Delta$. Meanwhile, a lower magnetic damping is also demanded since the energy required for switching the magnetization of the recording layer is proportional to the magnetic damping constant and $K_\mathrm{u}$ [1-4].

A CoFeB/MgO/CoFeB multilayer has been the most common MTJ structure used for the MRAMs owing to the high TMR ratio as well as the large interfacial PMA obtained by the solid-state epitaxy of CoFe(B)/MgO(001). [4-15] The magnetic and electrical transport properties of the CoFe(B)/MgO multilayer can be tailored by various approaches, such as changing CoFe(B) alloy compositions, [16, 17] and/or inserting an ultrathin film of Mg [18-20], Al [21, 22], or LiF [23] in between CoFe(B) and MgO. Another intriguing approach is to use Mg-*X*-O (*X*: transition metals) as a tunnel barrier. [24] For example, MTJs with Fe-substituted MgO (MgFeO) exhibited a clear PMA as well as high TMR ratio, accompanied by segregation of Fe from MgFeO onto both sides of the MgFeO barrier. [24] Although the segregation of Fe may play a key role on the magnetic and electrical transport properties at the CoFeB/MgFeO interface, details on the properties of segregated Fe such as the chemical states and magnetism have been remained as open issues.

In this work, we investigated the chemical states of Fe in the CoFeB/MgFeO multilayers by using X-ray photoelectron spectroscopy (XPS). The XPS measurement revealed that Fe in the MgFeO layer was ionized as $Fe^{2+}$, while the segregated Fe was reduced to metal. We also found that the segregated Fe exhibits remarkable ferromagnetism from magnetic measurements, and that can be used to enhance the PMA while reducing the magnetic damping. The use of atomic segregation from tunnel barrier will be a promising approach for tailoring the interfacial magnetic properties of MTJs to improve the performance of MRAMs.

*Experimental Methods*

Multilayers consisting of Ta(5)/Pt(5)/Ru(10)/TaB(10)/Pt(5)/Ta(3)/Mo(1)/CoFeB($t_{CoFeB}$)/MgO($t_{MgO}$)/Mg$_{40}$Fe$_{10}$O$_{50}$($t_{MgFeO}$) with a capping layer were deposited on Si/SiO$_2$ substrates (thicknesses in nm) by sputtering method. The deposition rates of MgO and MgFeO were calibrated by using X-ray reflectivity measurement. Those multilayers were annealed at $T_a$ of 300 – 400°C in a vacuum furnace without a magnetic field. XPS measurement was carried out by setting the detection angle of photoelectrons to be normal to the sample plane, and XPS depth profiles were collected by repeating Ar ion etching and the subsequent XPS measurement [Fig. 1(a)]. Magnetization curves of the multilayers were measured by using a vibrating sample magnetometer (VSM) under an in-plane magnetic field ($\mu_0 H$) up to 1.2 T. The dynamical magnetic property was investigated by using a vector network analyzer ferromagnetic resonance (VNA-FMR) spectrometer with field modulation. [25,26] FMR was excited under an out-of-plane $\mu_0 H$ up to 1 T.

*Experimental Results*

First, the XPS was used to investigate the chemical states of Fe both inside and outside the MgFeO layer. For the XPS measurement, we prepared a CoFeB(3)/MgFeO(10)/Ru(3) multilayer annealed at 300°C. Although it is expected that the Fe segregated from the MgFeO layer diffuse into both CoFeB and Ru layers, since the bottom CoFeB obscures the signal from segregated Fe, we shall discuss the chemical state of Fe at the MgFeO/Ru interface. Figures 1(b) – 1(d) show the typical Fe-$2p$ spectra obtained from the MgFeO/Ru interface, at the middle of MgFeO layer, and at the middle of CoFeB layer, respectively. In the CoFeB region, XPS peaks were observed at 720.1 and 707.0 eV, which are consistent with the metallic Fe $2p_{1/2}$ and $2p_{3/2}$ peaks, respectively. [27-29] On the other hand, Fe $2p_{1/2}$ and $2p_{3/2}$ peaks in the MgFeO region were observed at 723.5 and 709.8 eV, respectively, which coincide with the existence of $Fe^{2+}$ in FeO rather than the other Fe-oxides such as $Fe_3O_4$, and $Fe_2O_3$. [28,29] In fact, the spectral shape of MgFeO around the Fe $2p$ peaks resembles to that reported for FeO including satellite peaks. [28] Therefore, it is considered that the Fe inside the MgFeO layer was mostly ionized as $Fe^{2+}$, *i.e.,* Fe substituted the Mg sites of the MgO crystal. In contrast to the bulk MgFeO region, at the MgFeO/Ru interface, the Fe $2p$ peak was observed at around 706.9 eV as shown in Fig. 1(b). The shift of the Fe $2p$ peak to the low energy side is indicative of the reduction of Fe to the metallic state associated with the segregation from the MgFeO layer.

Then, we investigated the influence of the segregated Fe on the magnetic properties of the CoFeB/MgFeO multilayers. Figure 2(a) shows the CoFeB thickness ($t_{CoFeB}$) dependence of the saturation magnetization ($\mu_0 M_s t_{CoFeB}$) for the CoFeB/MgO and CoFeB/MgFeO multilayers annealed at 400°C. Linear relation between $\mu_0 M_s t_{CoFeB}$ and $t_{CoFeB}$ was obtained for both CoFeB/MgO and CoFeB/MgFeO multilayers. The slopes of $\mu_0 M_s t_{CoFeB}$ vs $t_{CoFeB}$ obtained from the linear fitting to the experimental results were almost the same for both structures, indicating that $\mu_0 M_s$ of CoFeB layer remain unchanged by changing MgO into MgFeO. On the other hand, the $\mu_0 M_s t_{CoFeB}$ values of CoFeB/MgFeO were found to be slightly larger than those of CoFeB/MgO in the whole range of $t_{CoFeB}$. Figure 2(b) shows annealing temperature dependence of the *x*-intercept of $\mu_0 M_s t_{CoFeB}$ vs $t_{CoFeB}$ plot obtained from the linear fitting, *i.e.*, the dead layer thickness ($t_{dead}$). $t_{dead}$ was found to be almost zero in the CoFeB/MgO multilayer regardless of the annealing temperature, while the CoFeB/MgFeO exhibited negative $t_{dead}$ for this annealing temperature range. The negative $t_{dead}$ in CoFeB/MgFeO multilayers suggests an increased effective magnetic layer due to the ferromagnetism of Fe segregated from the MgFeO layer.

Figure 3(a) shows in-plane magnetization curves of CoFeB(0.9)/MgO($t_{MgO}$)/MgFeO($t_{MgFeO}$) multilayers annealed at 300°C. $t_{MgO}$ and $t_{MgFeO}$ were varied while the total thickness of MgO/MgFeO was fixed to be 2.3 nm. Note that $t_{MgO}$ = 0 and 2.3 nm correspond to CoFeB/MgFeO and CoFeB/MgO structures, respectively. Although all these multilayers showed remarkable PMA, the CoFeB/MgO multilayer exhibited a rounded magnetization curve suggesting an inhomogeneity in the PMA along the film plane. For $t_{MgO} \leq 0.6$ nm, the multilayers exhibited typical hard-axis magnetization curves representing uniform rotation of magnetization, and the largest saturation field of around 0.3 T was obtained for the CoFeB(0.9)/MgO(0.4)/MgFeO(1.9) multilayer. Figures 3(b) and (c) show the $t_{MgO}$ dependence of $\mu_0 M_s t_{CoFeB}$ and the PMA energy density $K_u t_{CoFeB}$ obtained for the CoFeB/MgO/MgFeO multilayers, respectively. An enhanced $\mu_0 M_s t_{CoFeB}$ was observed only for $t_{MgO} < 0.6$ nm, *i.e.*, the segregation of ferromagnetic Fe from the MgFeO layer was suppressed by the insertion of MgO layer thicker than 0.6 nm. As for the PMA, more than 2-fold enhancement in $K_u t_{CoFeB}$ was obtained for the CoFeB/MgFeO multilayer compared with the CoFeB/MgO multilayer annealed at 300°C. $K_u t_{CoFeB}$ was further enhanced by inserting an ultrathin MgO layer in between the CoFeB and MgFeO layers; the highest $K_u t_{CoFeB}$ of 188 µJ/m$^2$ was obtained for $t_{MgO}$ = 0.4 nm. The enhanced $K_u t_{CoFeB}$ in the CoFeB/MgO/MgFeO multilayer can be attributed to improved (001)-orientation of MgFeO by insertion of the pure MgO seed layer combined with the segregation of ferromagnetic Fe from the MgFeO layer. The $K_u t_{CoFeB}$ value in the CoFeB(0.9)/MgO(0.4)/MgFeO(1.9) multilayer was maximized after annealing at 350°C, and $K_u t_{CoFeB}$ > 200 µJ/m$^2$ was maintained even after annealing at 400°C.

Finally, we carried out the VNA-FMR measurement to investigate the magnetic damping as well as the PMA in the multilayers. Figure 4(a) shows the resonance frequency $f_{FMR}$ as a function of out-of-plane $\mu_0 H$ for three different multilayers, namely, CoFeB (0.9)/MgO (2.3), CoFeB (0.9)/MgO (0.4)/MgFeO (1.9), and CoFeB(0.9)/MgFeO (2.3). The annealing temperature was 300°C. For a perpendicularly magnetized thin film, $f_{FMR}$ is proportional to $\mu_0 H$:

$$f_{FMR} = \frac{ge}{4\pi m}\mu_0(H - H_{k,eff}), \qquad (1)$$

where $g$, $e$, $m$, $\mu_0$, and $H_{k,eff}$ are Lande factor, charge of electron, mass of electron, permeability of vacuum, and effective anisotropy field, respectively. [25] The $\mu_0 H_{k,eff}$ values obtained from the linear extrapolation of $f_{FMR}$ vs $\mu_0 H$ plot were 20.8, 286.1, and 189.6 mT for the CoFeB/MgO, CoFeB/MgO/MgFeO, and CoFeB/MgFeO multilayers, respectively. These $\mu_0 H_{k,eff}$ values agree well with the saturation field obtained from the in-plane magnetization curves shown in Fig. 3(a).

Figure 4(b) shows the $f_{FMR}$ dependence of FMR spectral linewidth ($\mu_0 \Delta H$). It is clearly seen that the CoFeB/MgO/MgFeO, and CoFeB/MgFeO multilayers exhibited substantially smaller $\mu_0 \Delta H$ compared with the CoFeB/MgO multilayer. $\mu_0 \Delta H$ reflects the relaxation process of the magnetization precession and can be decomposed as the contributions of total magnetic damping $\alpha_{tot}$ and the inhomogeneous linewidth $\mu_0 \Delta H_0$ as follows.[26]

$$\mu_0 \Delta H = \frac{2h}{g\mu_B}\alpha_{tot}f_{FMR} + \mu_0 \Delta H_0, \qquad (2)$$

where $h$ and $\mu_B$ are Planck constant and Bohr magneton, respectively. $\alpha_{tot}$ includes the intrinsic magnetic damping and the interfacial spin relaxation due to the spin pumping effect, and $\mu_0 \Delta H_0$ is a measure of the spatial inhomogeneity in the anisotropy field [30]. The obtained values of $\alpha_{tot}$ and $\mu_0 \Delta H_0$ are summarized in Table I along with the $K_u t_{CoFeB}$ and $\mu_0 H_{k,eff}$ values. The use of MgFeO led to a remarkable enhancement in the PMA while reducing $\alpha_{tot}$ and $\mu_0 \Delta H_0$. Since this trend is also confirmed in the CoFeB/MgO/MgFeO multilayer with sufficiently thin MgO layer, we consider that the segregation of ferromagnetic Fe from the MgFeO layer plays a key role in the achievement of large

PMA as well as a low $\alpha_{\text{tot}}$ and $\mu_0\Delta H_0$.

*Conclusion*

In summary, we investigated the chemical state of Fe segregated from MgFeO and its influence on the magnetic properties of CoFeB/MgFeO multilayers. XPS measurement revealed the segregation of metallic Fe from the MgFeO layer in which the existence of $Fe^{2+}$ ions were confirmed. The segregated Fe exhibited ferromagnetism as evidenced by the negative $t_{\text{dead}}$ obtained from the magnetic measurement. Substantial enhancement in the PMA was obtained for the CoFeB/MgFeO multilayers while reducing magnetic damping in both $\alpha_{\text{tot}}$ and $\mu_0\Delta H_0$. Our experimental results demonstrate that the segregation of ferromagnetic Fe from the MgFeO layer can be beneficial for developing high-density and energy efficient MRAMs.

*Acknowledgments*

This study was funded in part by the New Energy and Industrial Technology Development Organization's (NEDO) Project (No. JPNP20017). XPS measurement was conducted with support of "Advanced Research Infrastructure for Materials and Nanotechnology in Japan (ARIM)" of the Ministry of Education, Culture, Sports, Science and Technology (MEXT). Grant Number JPMXP1222AT0309.

Table I. Magnetic properties of CoFeB(0.9)/MgO(2.3), CoFeB(0.9)/MgO(0.4)/MgFeO(1.9), and CoFeB(0.9)/MgFeO(2.3) multilayers (thickness in nm) annealed at 300°C. Perpendicular magnetic anisotropy $K_u t_{CoFeB}$ was estimated from magnetization curves, and the other properties of effective anisotropy fields $\mu_0 H_{k,eff}$, total magnetic damping $\alpha_{tot}$, and inhomogeneous line width $\mu_0 \Delta H_0$ were estimated from ferromagnetic resonance measurement.

| Sample | $K_u t_{CoFeB}$ (μJ/m$^2$) | $\mu_0 H_{k,eff}$ (mT) | $\alpha_{tot}$ | $\mu_0 \Delta H_0$ (mT) |
|---|---|---|---|---|
| CoFeB/MgO | 71±3 | 20.8±0.4 | 0.015±0.001 | 21.9±0.5 |
| CoFeB/MgO/MgFeO | 188±9 | 286.1±0.3 | 0.012±0.001 | 13.9±0.6 |
| CoFeB/MgFeO | 142±7 | 189.6±0.1 | 0.012±0.001 | 12.9±0.6 |

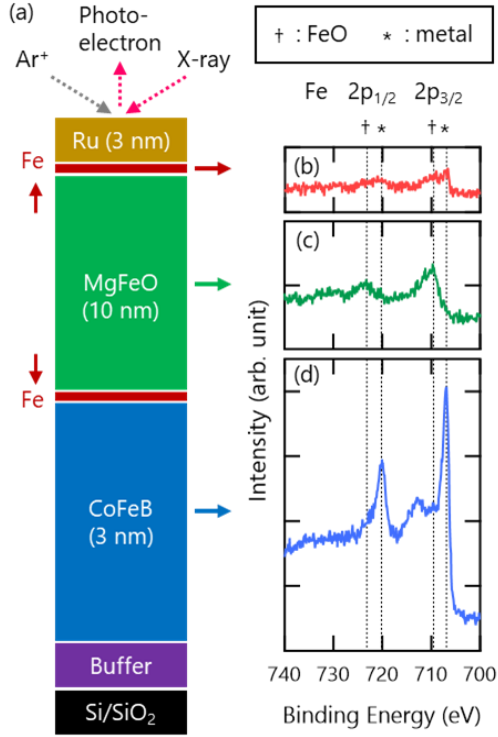

Figure 1. Depth profiles of X-ray photoelectron spectra (XPS) for Fe 2p in CoFeB/MgFeO/Ru multilayers annealed at 300°C. (a) Schematic illustration of sample structure and detection of XPS signals. XPS Fe 2p spectra from (b) the MgFeO/Ru interface, (c) at the middle of MgFeO layer, and (d) at the middle of CoFeB layer.

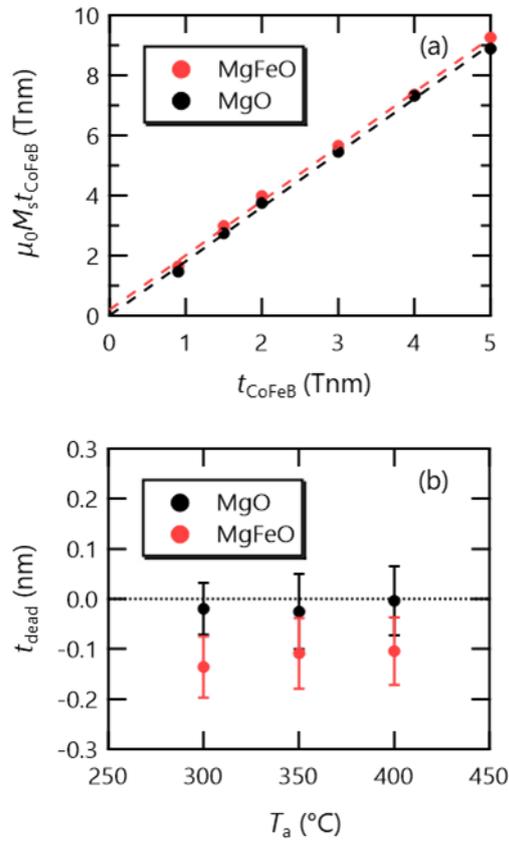

Figure 2. (a) Saturation magnetization ($\mu_0 M_s t_{CoFeB}$) as a function of CoFeB thickness ($t_{CoFeB}$) in CoFeB/MgO and CoFeB/MgFeO multilayers annealed at 400°C. (b) Annealing temperature ($T_a$) dependence of $t_{dead}$, i.e., the x-intercept of $\mu_0 M_s t_{CoFeB}$ vs $t_{CoFeB}$ plot obtained from the linear fitting.

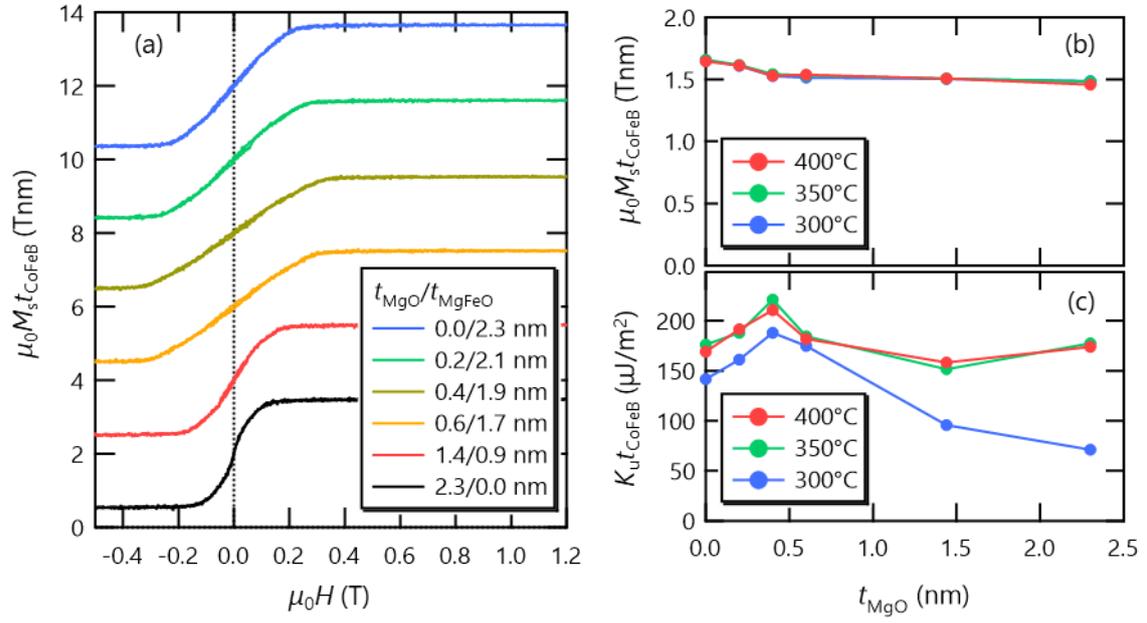

Figure 3. (a) Magnetization curves of CoFeB(0.9)/MgO($t_{MgO}$)/MgFeO($t_{MgFeO}$) multilayers annealed at 300°C (thickness in nm). $t_{MgO}$ dependence of (b) saturation magnetization ($\mu_0 M_s t_{CoFeB}$) and (c) perpendicular magnetic anisotropy ($K_u t_{CoFeB}$).

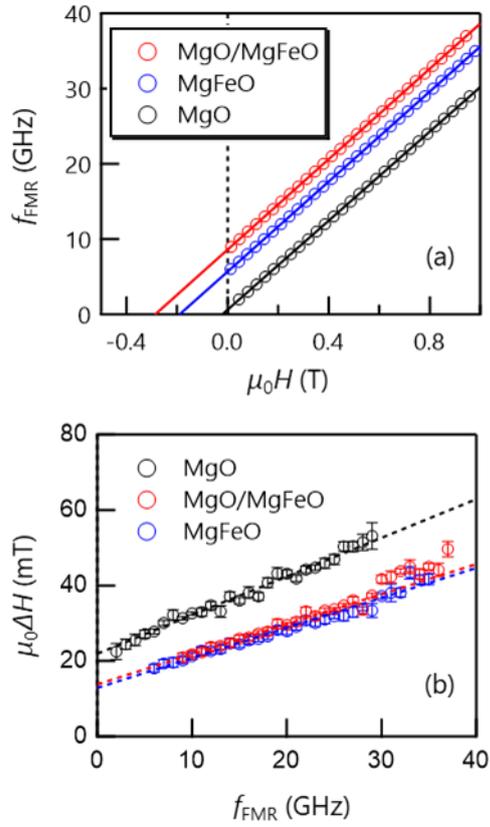

Figure 4. (a) Ferromagnetic resonance frequency ($f_{FMR}$) as a function of out-of-plane magnetic field ($\mu_0 H$) for CoFeB(0.9)/MgO(2.3), CoFeB(0.9)/MgO(0.4)/MgFeO(1.9), and CoFeB(0.9)/MgFeO(2.3) multilayers annealed at 300°C. (thickness in nm) (b) $f_{FMR}$ dependence of FMR spectral linewidth ($\mu_0 \Delta H$).

*References*

[1] B. Dieny, M. Chshiev, "Perpendicular magnetic anisotropy at transition metal/oxide interfaces and applications," Rev. Mod. Phys. **89**, 025008 (2017).

[2] B. Tudu, A. Tiwari, "Recent Developments in Perpendicular Magnetic Anisotropy Thin Films for Data Storage Applications," Vacuum **146**, 329 (2017).

[3] S. Ikegawa, F. B. Mancoff, J. Janesky, S. Aggarwal, "Magnetoresistive Random Access Memory: Present and Future," IEEE Trans. Electron Dev. **67**, 1407 (2020).

[4] S. Yuasa, K. Hono, G. Hu, D. C. Worledge, "Materials for spin-transfer-torque magnetoresistive random-access memory," MRS Bull. **43**, 352 (2018).

[5] S. S. P. Parkin, C. Kaiser, A. Panchula, P. M. Rice, B. Hughes, M. Samant, S.-H. Yang, "Giant tunneling magnetoresistance at room temperature with MgO(100) tunnel barriers," Nat. Mater. **3**, 862 (2004).

[6] S. Yuasa, T. Nagahama, A. Fukushima, Y. Suzuki, K. Ando, "Giant room-temperature magnetoresistance in single-crystal Fe/MgO/Fe magnetic tunnel junctions," Nat. Mater. **3**, 868 (2004).

[7] D. D. Djayaprawira, K. Tsunekawa, M. Nagai, H. Maehara, S. Yamagata, N. Watanabe, S. Yuasa, Y. Suzuki, K. Ando, Appl. Phys. Lett. **86**, 092502 (2005).

[8] S. Yakata, H. Kubota, Y. Suzuki, K. Yakushiji, A. Fukushima, S. Yuasa, K. Ando, "Influence of perpendicular magnetic anisotropy on spin-transfer switching current in CoFeB/MgO/CoFeB magnetic tunnel junctions," J. Appl. Phys. **105**, 07D131 (2009).

[9] S. Ikeda, K. Miura, H. Yamamoto, K. Mizunuma, H. D. Gan, M. Endo, S. Kanai, J. Hayakawa, F. Matsukura, H. Ohno, "A perpendicular-anisotropy CoFeB-MgO magnetic tunnel junction," Nat. Mater. **9**, 721 (2010).

[10] H. Sato, M. Yamanouchi, S. Ikeda, S. Fukami, F. Matsukura, H. Ohno, "Perpendicular-anisotropy

CoFeB-MgO magnetic tunnel junctions with a MgO/CoFeB/Ta/CoFeB/MgO recording structure," Appl. Phys. Lett. **101**, 022414 (2012).

[11] H. Cheng, J. Chen, S. Peng, B. Zhang, Z. Wang, D. Zhu, K. Shi, S. Eimer, X. Wang, Z. Guo, Y. Xu, D. Xiong, K. Cao, W. Zhao, "Giant Perpendicular Magnetic Anisotropy in Mo-Based double-Interface Free Layer Structure for Advanced Magnetic Tunnel Junctions," Adv. Electron. Mater. **6**, 2000271, (2020).

[12] H. X. Yang, M. Chshiev, B. Dieny, J. H. Lee, A. Manchon, K. H. Shin, "First-principles investigation of the very large perpendicular magnetic anisotropy at Fe|MgO and Co|MgO interfaces," Phys. Rev. B **84**, 054401 (2011).

[13] A. Hallal, H. X. Yang, B. Dieny, M. Chshiev, "Anatomy of perpendicular magnetic anisotropy in Fe/MgO magnetic tunnel junctions: First-principles insight," Phys. Rev. B **88**, 184423 (2013).

[14] G. Cai, Z. Wu, f. Guo, Y. Wu, H. Li, Q. Liu, M. Fu, T. Chen, J. Kang, "First-principles calculations of perpendicular magnetic anisotropy in $Fe_{1-x}Co_x$/MgO(001) thin films," Nanoscale Res. Lett. **10**, 126 (2015).

[15] S. Nazir, K. Yang, "Elucidate interfacial disorder effects on the perpendicular magnetic anisotropy at Fe/MgO heterostructure from first-principles calculations," J. Phys.: Condens. Matter **34**, 214009 (2022).

[16] M. Kodzuka, T. Ohkubo, K. Hono, S. Ikeda, H. D. Gan, H. Ohno, "Effects of boron composition on tunneling magnetoresistance ratio and microstructure of CoFeB/MgO/CoFeB pseudo-spin-valve magnetic tunnel junctions," J. Appl. Phys. **111**, 043913 (2012).

[17] S. Andrieu, L. Calmels, T. Hauet, F. Bonell, P. Le Fevre, F. Bertran, "Spectroscopic and transport studies of $Co_xFe_{1-x}$/MgO(001)-based magnetic tunnel junctions," Phys. Rev. B **90**, 214406 (2014).

[18] Q. L. Ma, S. Iihama, T. Kubota, X. M. Zhang, S. Mizukami, Y. Ando, T. Miyazaki, "Effect of Mg interlayer on perpendicular magnetic anisotropy of CoFeB films in MgO/Mg/CoFeB/Ta structure,"

Appl. Phys. Lett. **101**, 122414 (2012).

[19] X. Li, K. Fitzell, D. Wu, C. T. Karaba, A. Buditama, G. Yu, K. L. Wong, N. Altieri, C. Grezes, N. Kioussis, S. Tolbert, Z. Zhang, J. P. Chang, P. K. Amiri, K. L. Wang, "Enhancement of voltage-controlled magnetic anisotropy through precise control of Mg insertion thickness at CoFeB/MgO interface," Appl. Phys. Lett. **110**, 052401 (2017).

[20] C. A. Pandey, X. Li, H. Wang, H. Wang, Y. Liu, H. Zhao, Q. Yang, T. Nie, W. Zhao, "Enhancing perpendicular magnetic anisotropy through dead layer reduction utilizing precise control of Mg insertion," J. Magn. Magn. Mater. **511**, 166956 (2020).

[21] K. Yakushiji, A. Sugihara, T. Nakano, S. Yuasa, "Fully epitaxial magnetic tunnel junction on a silicon wafer," Appl. Phys. Lett. **115**, 202403 (2019).

[22] T. Nozaki, T. Nozaki, H. Onoda, H. Nakayama, T. Ichinose, T. Yamamoto, M. Konoto, S. Yuasa, "Precise interface engineering using a post-oxidized ultrathin MgAl layer for the voltage=controlled magnetic anisotropy," APL Mater. **10**, 081103 (2022).

[23] T. Nozaki, T. Nozaki, T. Yamamoto, M. Konoto, A. Sugihara, K. Yakushiji, H. Kubota, A. Fukushima, S. Yuasa, "Enhancing the interfacial perpendicular magnetic anisotropy and tunnel magnetoresistance by inserting an ultrathin LiF layer at an Fe/MgO interface," NPG Asia Mater. **14**, 5 (2022).

[24] K. Yakushiji, E. Kitagawa, T. Ochiai, H. Kubota, N. Shimomura, J. Ito, H. Yoda, S. Yuasa, "Fabrication of Mg-X-O (X = Fe, Co, Ni, Cr, Mn, Ti, V, and Zn) barriers for magnetic tunnel junctions," AIP Adv. **8**, 055905 (2018).

[25] S. Tamaru, T. Yamamoto, T. Onuma, N. Kikuchi, S. Okamoto, "Development of a high-sensitivity VNA-FMR spectrometer with field modulation detection and its application to magnetic characterization," IEEJ Trans. Fund. Mater. **141**, 295 (2021).

[26] A. Sugihara, T. Ichinose, S. Tamaru, T. Yamamoto, M. Konoto, T. Nozaki, S. Yuasa, "Low

magnetic damping in an ultrathin CoFeB layer deposited on a 300 mm diameter wafer at cryogenic temperature," Appl. Phys. Express **16**, 023003 (2023).

[27] P. Mills, J. L. Sullivan, "A study of the core level electrons in iron and its three oxides by means of x-ray photoelectron spectroscopy," J. Phys. D: Appl. Phys. **16**, 723 (1983).

[28] T. Yamashita, P. Hayes, "Analysis of XPS spectra of $Fe^{2+}$ and $Fe^{3+}$ ions in oxide materials," Appl. Surf. Sci. **254**, 2441 (2008).

[29] Y. Liu, J. Zhang, S. Wang, S. Jiang, Q. Liu, X. Li, Z. Wu, G. Yu, "Ru Catalyst-Induced Perpendicular Magnetic Anisotropy in MgO/CoFeB/Ta/MgO Multilayered Films," ACS Appl. Mater. Interfaces **7**, 26643 (2015).

[30] A. Okada, S. He, B. Gu, S. Kanai, A. Soumyanarayanan, S. T. Lim, M. Tran, M. Mori, S. Maekawa, F. Matsukura, H. Ohno, C. Panagopoulos, "Magnetization dynamics and its scattering mechanism in thin CoFeB films with interfacial anisotropy," Proc. Natl. Acad. Sci. USA **114**, 3815 (2017).